\documentclass[floatfix,twocolumn,prl,amsmath]{revtex4-2}
\usepackage{graphicx}
\usepackage{bm}
\usepackage{amssymb}
\usepackage{dcolumn}
\usepackage{subfigure}
\usepackage{threeparttable}
\usepackage{multirow}
\usepackage{mathrsfs}
\DeclareMathAlphabet{\mathscrbf}{OMS}{mdugm}{b}{n}
\usepackage{color}
\begin{document}
\newcommand {\grbf}[1]{\rm{\boldmath $ #1 $}}
\newcommand{\vn}[1]{{\boldsymbol{#1}}}
\newcommand{\vht}[1]{{\boldsymbol{#1}}}
\newcommand{\matn}[1]{{\bf{#1}}}
\newcommand{\matnht}[1]{{\boldsymbol{#1}}}
\newcommand{\bege}{\begin{equation}}
\newcommand{\ee}{\end{equation}}
\newcommand{\bal}{\begin{aligned}}
\newcommand{\defbar}{\overline}
\newcommand{\SM}{\scriptstyle}
\newcommand{\gretke}{G_{\vn{k} }^{\rm R}(\mathcal{E})}
\newcommand{\gret}{G^{\rm R}}
\newcommand{\gadv}{G^{\rm A}}
\newcommand{\gmat}{G^{\rm M}}
\newcommand{\gles}{G^{<}}
\newcommand{\ghat}{\hat{G}}
\newcommand{\sigmahat}{\hat{\Sigma}}
\newcommand{\glesone}{G^{<,{\rm I}}}
\newcommand{\glestwo}{G^{<,{\rm II}}}
\newcommand{\sigmaret}{\Sigma^{\rm R}}
\newcommand{\sigmales}{\Sigma^{<}}
\newcommand{\sigmalesone}{\Sigma^{<,{\rm I}}}
\newcommand{\sigmalestwo}{\Sigma^{<,{\rm II}}}
\newcommand{\sigmalesthree}{\Sigma^{<,{\rm III}}}
\newcommand{\polarivec}{\boldsymbol{\epsilon}}
\newcommand{\sigmaadv}{\Sigma^{A}}
\newcommand{\Bxc}{\Omega}
\newcommand{\mubo}{\mu_{\rm B}^{\phantom{B}}}
\newcommand{\rmd}{{\rm d}}
\newcommand{\rme}{{\rm e}}
\newcommand{\crea}[1]{{c_{#1}^{\dagger}}}
\newcommand{\annihi}[1]{{c_{#1}^{\phantom{\dagger}}}}
\newcommand{\intkspa}{\int\!\!\frac{\rmd^d k}{(2\pi)^d}}
\newcommand{\eal}{\end{aligned}}
\newcommand{\udot}{\overset{.}{u}}
\newcommand{\exponential}[1]{{\exp(#1)}}
\newcommand{\phandot}[1]{\overset{\phantom{.}}{#1}}
\newcommand{\phandag}{\phantom{\dagger}}
\newcommand{\Trace}{\text{Tr}}
\setcounter{secnumdepth}{2}
\title{
Laser-induced torques in spin spirals
}
\author{Frank Freimuth$^{1,2}$}
\email[Corresp.~author:~]{f.freimuth@fz-juelich.de}
\author{Stefan Bl\"ugel$^{1}$}
\author{Yuriy Mokrousov$^{1,2}$}
\affiliation{$^1$Peter Gr\"unberg Institut and Institute for Advanced Simulation,
Forschungszentrum J\"ulich and JARA, 52425 J\"ulich, Germany}
\affiliation{$^2$ Institute of Physics, Johannes Gutenberg University Mainz, 55099 Mainz, Germany
}
\begin{abstract}
We investigate laser-induced torques in magnetically non-collinear
ferromagnets with a spin-spiral magnetic structure using 
\textit{ab-initio} calculations. 
Since spin-spirals
may be used to approximate the magnetization
gradients locally in domain walls and skyrmions, our method may
be used to obtain the laser-induced torques 
in such objects from a multiscale approach.
Employing the generalized Bloch-theorem we obtain the electronic 
structure computationally efficiently.
We employ our method to assess the laser-induced torques
in bcc Fe, hcp Co, and L$_{1}0$ FePt when a spin-spiral magnetic
structure is imposed.
We find that the laser-induced torques in these magnetically
noncollinear systems
may be orders of magnitude larger than those
 in the corresponding magnetically collinear systems and that they exist
both for linearly and circularly polarized light.
This result suggests that laser-induced torques driven by
noncollinear magnetic order or by magnetic fluctuations
may contribute significantly to processes in ultrafast magnetism.
\end{abstract}

\maketitle
\section{Introduction}
Femtosecond laser-pulses exert effective magnetic
fields on the magnetization 
in collinear ferromagnets, which may be used to 
tilt the magnetization and to excite
magnetization 
dynamics~\cite{Huisman_2016,Rabi_CoFeB,optical_driven_magnetization_dynamics_choi_2017}.
These effective magnetic fields have been ascribed to
the inverse Faraday effect (IFE) and to the optical
spin transfer torque
(OSTT)~\cite{Kimel_ultrafast_control_magnetization,nemec_ostt,lasintor,Li_Haney_optical_spin_transfer}.
The IFE is a key ingredient in several theoretical explanations
of magnetization reversal in ferromagnetic thin 
films~\cite{Lambert_all_optical_control,Mag_switching_Berritta}.

In spintronics, the spin-orbit torque~\cite{rmp_sot} 
requires spin-orbit interaction (SOI),
while the spin-transfer torque~\cite{RALPH20081190} does not.
The reason is that in collinear ferromagnets the angular
momentum can only be transferred to the lattice, which
requires SOI, while in non-collinear magnets, spin-valves
and magnetic tunnel junctions the angular momentum 
is transferred between different magnetization directions
and between different magnetic layers, which 
does not require SOI.
IFE and OSTT have been studied mostly in
collinear magnets and they require SOI.  

This comparison between spintronics and 
laser-induced ultrafast magnetism therefore poses the 
question if there are 
additional laser-induced
torques in non-collinear magnets or in spin-valves
that arise from 
different mechanisms than the IFE and the OSTT.
Indeed, laser-pulses may distort domain walls and excite magnetization
dynamics in them~\cite{2020arXiv200203971K,2020arXiv200708583L}. 
Additionally, laser-pulses excite spin currents in spin-valves,
which generate spin-transfer torques when they flow
between different magnetic layers~\cite{ultrafast_stt_laser,choi_thermal_stt}.
These laser-induced spin-transfer torques resemble the
Slonczewski spin-transfer torque in spintronics. 
However, in spintronics a second type of torque 
is known, the so-called non-adiabatic 
torque~\cite{RALPH20081190,spin_torques_kohno_tatara_shibata,functional_keldysh_spin_torques,nonadiabatic_stt_garate_macdonald,intrinsic_torque_no_soi}.  
Therefore, one may expect that not only the Slonczewski torque,
but also the non-adiabatic torque should have a laser-induced
counterpart in ultrafast magnetism. We will confirm
this expectation in this paper.

Strong femtosecond laser-pulses
do not only generate effective magnetic fields,
but they also trigger ultrafast demagnetization.
There are many indications that ultrafast demagnetization
in transition metal collinear ferromagnets 
is not dominated by collapsed exchange
but rather by collective
excitations~\cite{ufd_pastor,collapsed_vs_collective,longitex}.
Since the magnetization is non-collinear in the presence
of collective excitations, one may pose the question
if the
laser-induced torques that arise from this non-collinearity
might contribute to the ultrafast demagnetization itself. 
Moreover, in order to describe the processes involved in
ultrafast magnetism at room temperature properly
it is crucial to take the initial thermal fluctuations of the magnetization
into
account~\cite{initial_magnetic_disorder,afm_sumit}. 
Within the limitations of the frozen magnon approximation
our results on spin spirals may also be used to estimate the laser-induced
torques on magnons, which therefore provides a valuable asset to
understand the torques active in ultrafast magnetism.

Several recent works have added an ultrafast-magnetism perspective to
magnetically noncollinear objects such as
skyrmions~\cite{PhysRevLett.110.177205,creation_skymion_laser,2020heating_path_nucleation_skyrmion}
and
domain walls~\cite{2020arXiv200203971K,2020arXiv200708583L}.
In order to apply our results for homogeneous spin spirals
to such inhomogeneous objects one may locally approximate the
magnetization gradients by spin spirals and use a multiscale 
approach~\cite{PhysRevLett.108.077201}.

This paper is structured as follows.
In Sec.~\ref{sec_formalism} we present our theory
and computational formalism
of laser-induced torques in spin spirals.
In Sec.~\ref{sec_lasintospira}
we introduce basic notations. 
In Sec.~\ref{sec_symmetry}
we discuss symmetry properties of
laser-induced torques.
In Sec.~\ref{sec_diffmech}
we explain key differences between 
laser-induced torques on spin spirals
and the current-induced torques known
from spintronics. 
In Sec.~\ref{sec_gradi} we develop a simple model
useful to understand laser-induced torques on spin spirals.
In Sec.~\ref{sec_gauge_field_approach} we describe how key properties of 
the laser-induced torques may be understood within the 
gauge-field approach.
In Sec.~\ref{sec_formali} we describe our computational
method.
In Sec.~\ref{sec_results} we discuss our
\textit{ab-initio} results on the laser-induced
torques in bcc Fe, hcp Co, and L$_{1}$0 FePt.
This paper ends with a summary in
 Sec.~\ref{sec_conclusion}.
\section{Theory}
\label{sec_formalism}
\subsection{Laser-induced torques on spin-spirals}
\label{sec_lasintospira}
The magnetization direction $\hat{\vn{M}}$ of spin spirals may be written as
\bege\label{eq_spin_spiral}
\hat{\vn{M}}(\vn{r})=\mathcal{R}(\alpha,\beta)
\begin{pmatrix}
\sin(\theta)\cos(\vn{q}\cdot \vn{r} +\phi)\\
\sin(\theta)\sin(\vn{q}\cdot \vn{r}+\phi)\\
\cos(\theta) 
\end{pmatrix},
\ee
where $\vn{q}$ is the spin-spiral wave-vector, $\theta$ is the cone
angle of the spin-spiral, and $\mathcal{R}(\alpha,\beta)$ is a proper orthogonal rotation matrix
parameterized by the two Euler angles $\alpha$ and $\beta$.
When $\alpha=\beta=0$ we have $\mathcal{R}(0,0)=1$ and
Eq.~\eqref{eq_spin_spiral} describes a helical spin spiral when
$\theta=90^{\circ}$ and when $\vn{q}$ points into the $z$-direction.
When $\vn{q}$ lies in the $xy$ plane it describes a cycloidal spiral.
Non-zero Euler angles are needed in Eq.~\eqref{eq_spin_spiral} 
to describe e.g.\ a helical spin
spiral propagating in $x$ direction, or a cycloidal spin spiral
propagating in $z$ direction.

Since torques on the magnetization are perpendicular to it, any torque
on the magnetization of a spin spiral may be expressed as
\bege\label{eq_torque_on_spiral}
\vn{T}(\vn{r})=
\mathcal{R}(\alpha,\beta)
\left[
\hat{\vn{e}}_{\theta}(\vn{r}) T_{\theta}+\hat{\vn{e}}_{\phi}(\vn{r})
T_{\phi}
\right], 
\ee
where the unit vectors $\hat{\vn{e}}_{\theta}(\vn{r})$
and $\hat{\vn{e}}_{\phi}(\vn{r})$ are given by
\bege
\begin{aligned}
\hat{\vn{e}}_{\theta}(\vn{r})=
\begin{pmatrix}
\cos(\theta)\cos(\vn{q}\cdot \vn{r} +\phi)\\
\cos(\theta)\sin(\vn{q}\cdot \vn{r}+\phi)\\
-\sin(\theta) 
\end{pmatrix}
\end{aligned}
\ee
and
\bege
\hat{\vn{e}}_{\phi}(\vn{r})=
\begin{pmatrix}
-\sin(\vn{q}\cdot \vn{r} +\phi)\\
\cos(\vn{q}\cdot \vn{r}+\phi)\\
0
\end{pmatrix},
\ee
respectively.

In spintronics, current-induced contributions to the torque
 $T_{\phi}$ are referred to as the \textit{adiabatic} torque
while current-induced contributions to $T_{\theta}$ are referred to as the \textit{non-adiabatic} 
torque~\cite{RALPH20081190}. It is nowadays agreed that the
terms \textit{adiabatic}
and \textit{non-adiabatic} are not optimal to describe the mechanisms
involved in these current-induced torques. However, these
two terms are well established to distinguish the two possible
directions of the current-induced torque.
When a torque is generated by the application of
laser-light, this torque may be decomposed as well into
the two components $T_{\phi}$ and $T_{\theta}$, 
which we denote therefore 
\textit{laser-induced adiabatic
torque} and \textit{laser-induced non-adiabatic
torque}, respectively. However, while borrowing these terms
from spintronics we do not suggest that the microscopic origin
of laser-induced torques is in any way similar to the microscopic
origin of current-induced torques. The analogies between the two,
which suggest to use the terms \textit{adiabatic}
and \textit{non-adiabatic}  for both effects, are only in the 
geometry, i.e., in the directions of these two components, and in the
effect of the torques on the magnetization dynamics.
The difference in mechanisms is explored in Sec.~\ref{sec_diffmech} 
and Sec.~\ref{sec_gradi}
below.

\subsection{Symmetry of laser-induced torques}
\label{sec_symmetry}
\begin{figure}
\includegraphics[width=\linewidth,trim=11.5cm 20cm 1cm 1.5cm, clip]{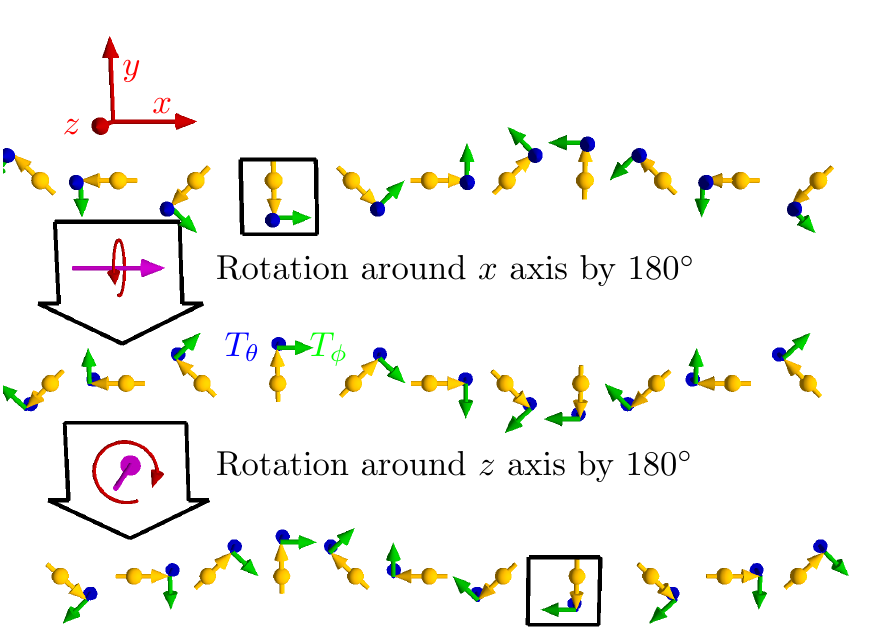}
\caption{\label{symmetry_argument1} Two consecutive 180$^{\circ}$
rotations around the $x$ and $z$ axes affect a flat spin spiral
(yellow arrows)
in the same way as a net translation of the spin spiral along the $x$
direction
does. However, these two consecutive rotations flip the signs of $T_{\theta}$
and
$T_{\phi}$ (compare the torques e.g.\ in the two square boxes). Consequently, symmetry requires the
laser-induced
torques to vanish in flat spin spirals.
}
\end{figure}

Consider a flat spin-spiral in the $xy$ plane, i.e.,
$\theta=90^{\circ}$, $\mathcal{R}=1$,
and $\vn{q}$ along the $x$ direction in Eq.~\eqref{eq_spin_spiral}.
Two subsequent rotations of the spin-spiral firstly by $180^{\circ}$
around the $x$ axis and secondly by $180^{\circ}$
around the $z$ axis lead to a simple translation
of the entire spin-spiral (see Fig.~\ref{symmetry_argument1}). 
However, the application of these
rotations to the torque reverses the sign of the torque.
Therefore, the laser-induced torques 
vanish in this case.
Generally, for flat spin-spirals, i.e., when $\theta=90^{\circ}$, the
laser-induced torques vanish.
Note that in this symmetry argument we do not consider SOI, because in
this 
work we consider laser-induced
torques that arise from the noncollinear magnetic order only.
In the presence of SOI the above symmetry argument does not hold
because
the two rotations are not allowed by symmetry.

\begin{figure}
\includegraphics[width=\linewidth,trim=11.6cm 20cm 1cm 1.5cm, clip]{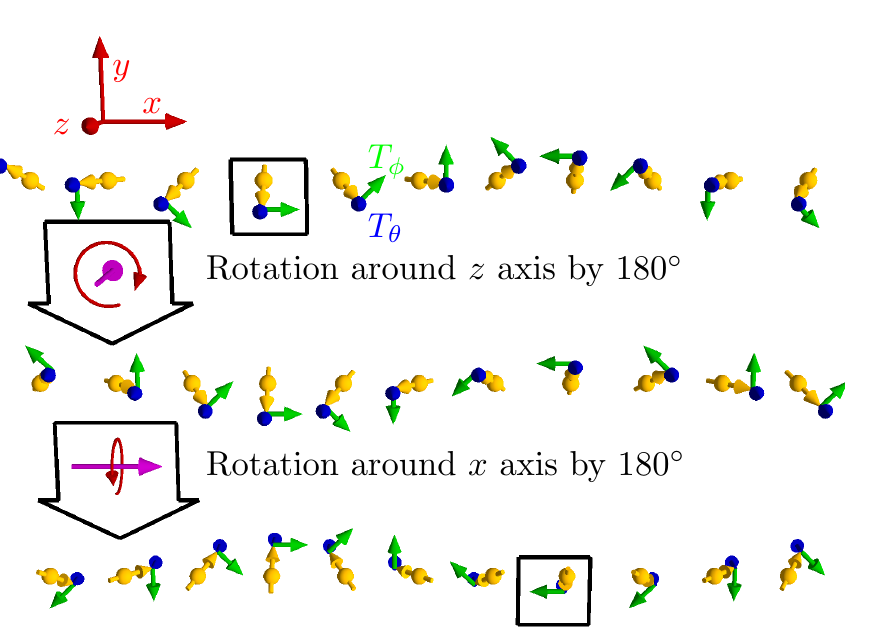}
\caption{\label{symmetry_argument2} 
A first rotation around the $z$ axis by 180$^{\circ}$ flips the sign of the
$q$ vector of the spin spiral, but does not affect the
torques. Consequently,
laser-induced torques are even in $q$. A second rotation around the
$x$ axis by 180$^{\circ}$ flips the sign of the torques (compare
e.g.\ the torques in the two square boxes). Consequently, the torques
satisfy $\vn{T}(\theta)=-\vn{T}(180^{\circ}-\theta)$.
}
\end{figure}

Consider a Neel-like spiral with $q$-vector in $x$ direction, cone
angle $\theta$, and  $\mathcal{R}=1$ in Eq.~\eqref{eq_spin_spiral}
(see Fig.~\ref{symmetry_argument2}).
When we rotate the spiral around the $z$ axis by 180$^{\circ}$ the 
$q$-vector changes sign, but the torque is not modified.
Consequently, laser-induced torques are even in $q$.
However, when we rotate next around the $x$ axis by 180$^{\circ}$
the torque changes sign. 
Therefore, $\vn{T}(\theta)=-\vn{T}(180^{\circ}-\theta)$.
This suggests that the dependence of the torques on $\theta$
may be described by $\propto \sin(2\theta)$ at the leading order.
As a special case it follows that the torque vanishes 
when $\theta=90^{\circ}$, consistent with the result above.

\subsection{Current-induced vs.\ laser-induced torques on
  spin-spirals: Different mechanisms}
\label{sec_diffmech}

While we borrowed the terms \textit{adiabatic}
and \textit{non-adiabatic} from spintronics in order to
distinguish the two components of the laser-induced
torques, the mechanisms responsible for current-induced
torques on spin spirals are quite different from those generating the
laser-induced torques. 
In this section we explore some of these
differences.

When an electric current propagates along a spin-spiral
the spin current is given by
\bege\label{eq_adiabatic_spin_current}
\vn{Q}_{i}(\vn{r})=\frac{\hbar}{2 e}
P J_{i} \hat{\vn{M}}(\vn{r}).
\ee
Here, the vector $\vn{Q}_{i}(\vn{r})$ describes the spin
current density flowing along the $i$-th cartesian direction.
This vector is parallel to the orientation of the spin polarization of
this spin current. $J_{i}$ is the electric current density
along the $i$-th cartesian direction and
\bege
P=\frac{
\sigma_{\uparrow}
-
\sigma_{\downarrow}
}{
\sigma_{\uparrow}
+
\sigma_{\downarrow}
}
\ee
is its polarization. $\sigma_{\uparrow}$ and $\sigma_{\downarrow}$ are the respective
contributions of the minority and majority electrons
to the electrical conductivity.
The resulting torque is given by
\bege\label{eq_current_induced_adiabatic_torque}
\vn{T}^{\rm adia}(\vn{r})=\sum_{i}\frac{\partial\vn{Q}_{i}}{\partial r_{i}}.
\ee
In Eq.~\ref{eq_adiabatic_spin_current} we assumed that the electron
spin
follows the local magnetization direction adiabatically. Therefore,
the torque in Eq.~\eqref{eq_current_induced_adiabatic_torque}
is called the adiabatic torque.

When a laser-pulse is applied to a homogeneous spin-spiral 
in a centrosymmetric crystal inversion symmetry does not allow
any spin current to be generated by the laser-pulse at the second
order of the electric field of the laser light. Of course, if the
laser
spot has a finite size, spin current will flow out of the illuminated
region, but we consider here the situation where the entire spiral
is homogeneously illuminated by the pulse. The absence of spin
currents in homogeneously illuminated spin spirals shows that
the microscopic mechanisms of laser-induced torques have to be
different than those of current-induced torques in spin spirals. 
Certainly, there are several experiments where laser pulses
excite superdiffusive spin currents, which  exert torques
on magnets~\cite{ultrafast_stt_laser}. However, in these experiments
two magnets are separated by a nonmagnetic spacer, which
is a geometry different to the one of a homogeneous spin spiral
that we consider in this work.

Since the spin current cannot explain the laser-induced torque in
homogeneous spin spirals, we develop a simple model in the
following section Sec.~\ref{sec_gradi}. 

\subsection{Gradient expansion}
\label{sec_gradi}
The Kohn-Sham Hamiltonian of a magnetic system may be 
written as
\bege\label{eq_hamil}
H(\vn{r})=H_{0}(\vn{r})-\vn{m}\cdot\hat{\vn{M}}(\vn{r})\Omega^{\rm xc}(\vn{r}),
\ee
where the first term, $H_{0}(\vn{r})$, contains kinetic energy and scalar potential,
while the exchange interaction is described by the second term. Here,
$\Omega^{\rm xc}(\vn{r})$ is the exchange field, i.e., the
difference between the potentials of majority and
minority electrons $\Omega^{\rm xc}(\vn{r})=\frac{1}{2\mu_{\rm
    B}}\left(V^{\rm eff}_{\rm minority}(\vn{r})-V^{\rm eff}_{\rm
    majority}(\vn{r}) \right)$, $\vn{m}=-\mu_{\rm B}\vht{\sigma}$,
$\mu_{\rm B}$  is the Bohr magneton 
and $\vht{\sigma}=(\sigma_x,\sigma_y,\sigma_z)^{\rm T}$
is the vector of Pauli spin matrices.  

In general the magnetization
of a spin spiral Eq.~\eqref{eq_spin_spiral} breaks the translational invariance of
the crystal lattice.
In order to obtain at a position $\vn{r}_{0}$ a local Hamiltonian that is consistent with the
crystal lattice translational symmetries one may expand the exchange
interaction in $H(\vn{r})$
around $\vn{r}_{0}$.
The expansion of the Hamiltonian around the position $\vn{r}_{0}$
is given by
\bege
\begin{aligned}
&H(\vn{r})=H_{0}(\vn{r}) -\vn{m}\cdot\hat{\vn{M}}(\vn{r}_{0})\Omega^{\rm xc}(\vn{r})\\
&-\Omega^{\rm xc}(\vn{r})\frac{\partial \{\vn{m}\cdot \hat{\vn{M}}(\vn{r}_{0})\} }{\partial \vn{r}_{0}}
\cdot[\vn{r}-\vn{r}_{0}]\\
&-\frac{1}{2}\Omega^{\rm xc}(\vn{r})
\sum_{ij}
\frac{\partial^2 \{
\vn{m}\cdot \hat{\vn{M}}(\vn{r}_{0})
\}
}{\partial r_{0,i} \partial r_{0,j}}
[r_{i}-r_{0,i}][r_{j}-r_{0,j}]+\cdots .
\end{aligned}
\ee
Consequently, the local Hamiltonian at $\vn{r}_{0}$ is
\bege
\begin{aligned}
&\langle H (\vn{r})\rangle \simeq H_{0}(\vn{r}) -\vn{m}\cdot\hat{\vn{M}}(\vn{r}_{0})\Omega^{\rm xc}(\vn{r})\\
&-\frac{1}{2}\Omega^{\rm xc}(\vn{r})
\sum_{ij}
\frac{\partial^2 \{
\vn{m}\cdot \hat{\vn{M}}(\vn{r}_{0})
\}
}{\partial r_{0,i} \partial r_{0,j}}
\langle
[r_{i}-r_{0,i}][r_{j}-r_{0,j}]
\rangle,
\end{aligned}
\ee
because $\langle[\vn{r}-\vn{r}_{0}]\rangle=0$
in systems with inversion symmetry. Here,
$\langle \dots \rangle$ denotes a suitable averaging.

The second derivative of the magnetization direction 
is given by

\bege
\begin{aligned}
\frac{\partial^2 \hat{\vn{M}}(\vn{r}_{0})}{\partial r_{0,i}\partial r_{0,j}}
=&-q_{i} q_{j}
\sin\theta
\Biggl[
\sin\theta\hat{\vn{M}}(\vn{r}_{0})
+\\
&+
\cos\theta
\mathcal{R}(\alpha,\beta)
\hat{\vn{e}}_{\theta}(\vn{r}_{0})
\Biggr],
\end{aligned}
\ee
which leads to an effective magnetic field
perpendicular to the local magnetization:
\bege\label{eq_perp_loc_bfi}
\vn{B}_{\vn{q}}(\vn{r})=-b_{\vn{q}}\Omega^{\rm xc}(\vn{r})
\sin(2\theta)
\mathcal{R}(\alpha,\beta)
\hat{\vn{e}}_{\theta}(\vn{r}),
\ee 
where
\bege\label{eq_bq}
b_{\vn{q}}=\frac{1}{4}
\sum_{ij}
q_{i}q_{j}
\langle
[r_{i}-r_{0,i}][r_{j}-r_{0,j}]
\rangle.
\ee

Consequently, the local Hamiltonian
may be written as
\bege
\begin{aligned}
&\langle H (\vn{r})\rangle \simeq H_{0}(\vn{r}) -\vn{m}\cdot
\Biggl[
\hat{\vn{M}}(\vn{r}_{0})\Omega^{\rm xc}(\vn{r})
+\vn{B}_{\vn{q}}(\vn{r})
\Biggr].\\
\end{aligned}
\ee
We assume that the application of a laser pulse
generates two torques, one in the direction of $\hat{\vn{M}}\times
\vn{B}_{\vn{q}}\propto \hat{\vn{e}}_{\phi}$
and a second one in the direction of
$\hat{\vn{M}}\times[\hat{\vn{M}}\times
\vn{B}_{\vn{q}}]\propto \hat{\vn{e}}_{\theta}$.
According to Eq.~\eqref{eq_perp_loc_bfi}
these torques are proportional to $\sin(2\theta)$, which is
consistent with the symmetry analysis in 
Sec.~\ref{sec_symmetry}. According to 
Eq.~\eqref{eq_bq} these torques are even in $\vn{q}$,
which is consistent with the
symmetry analysis in Sec.~\ref{sec_symmetry} 
and also with our \textit{ab-initio} results
in Sec.~\ref{sec_results}.
Thus, the assumption that the interaction of laser-excited
electrons with the effective magnetic field $\vn{B}_{\vn{q}}(\vn{r})$
leads to the laser-induced torques predicts a dependence 
on $\vn{q}$ and $\theta$ that agrees to the \textit{ab-initio}
results. Clearly, $\vn{B}_{\vn{q}}(\vn{r})$ exists even without any
applied laser-pulse. However, the expectation 
value $\langle[r_{i}-r_{0,i}][r_{j}-r_{0,j}]\rangle$ in Eq.~\eqref{eq_bq}
is state dependent, and therefore $b_{\vn{q}}$ changes when a
laser pulse is applied. 

When the $q$-vector is equal to a primitive vector $\vn{b}$ of the reciprocal lattice
one picks up the phase $2\pi$ over the length of
a primitive lattice vector. Consequently,
$\vn{q}=\vn{b}$
describes the same magnetic structure as $\vn{q}=0$. Similarly,
$\vn{q}=\vn{b}/2$ describes an antiferromagnet, where neighboring
magnetic atoms exhibit antiparallel magnetic moments. In such a
collinear
antiferromagnetic configuration the laser-induced torques are zero.
Therefore, we expect the torques to increase first with increasing
$q$, to attain a maximum around $\vn{b}/4$ and to decrease
afterwards
until they are zero at $\vn{b}/2$.
Thus, the model developed in this section, which predicts that
the torques increase with increasing $q$, is expected to be valid only
for $q<\pi/(2a)$, where $a$ is the lattice contant.

\subsection{Gauge-field approach to spin spirals}
\label{sec_gauge_field_approach}
The effects of magnetic texture on conduction electrons often
resemble those of SOI. In fact, mathematical exact transformations
of magnetization gradients into an effective SOI have been derived
and exploited
in important model systems (see Ref.~\cite{TATARA2019208} for a
recent review).
These relations have been used not only for the discussion of effects
linear in the magnetization gradients but also for effects e.g.\
quadratic
in the magnetization gradients~\cite{Karashtin_Tartara_second_order_gauge_field}. 
One may argue that this equivalence between magnetic non-collinearity and
effective SOI explains why laser-induced torques exist in spin-spirals
even
without real SOI, while collinear ferromagnets exhibit non-zero
laser-induced
torques only in the presence of SOI: Instead of the real SOI it is the
effective
SOI due to the magnetic non-collinearity that generates these torques
in spin-spirals even without any real atomic SOI.
This argument has been used to predict an IFE
in topological magnetic structures even without SOI~\cite{topological_IFE}.

In this work we consider Fe, Co and FePt. In these materials the SOI
strength on the magnetic atom is of the order of 60~meV.
Using the gauge-field approach from Ref.~\cite{chigyromag} 
we estimate that non-collinearity produces an effective SOI of the
order of magnitude of
\bege
\frac{\hbar^2 qk}{2m}\approx 1.5{\rm eV},
\ee 
where we set $q=k=2\pi/(10$\AA).
This is larger than the real SOI by a factor of 25. Consequently,
we expect the laser-induced torques from non-collinearity to be larger than those
from real SOI in these materials.

In sections Sec.~\ref{sec_symmetry} and Sec.~\ref{sec_gradi} 
we have shown that the laser-induced torques in spin spirals
are expected to be even in the $q$-vector and to exhibit the
angular dependence $\propto \sin (2\theta)$. In the following
we show how these dependences may be understood within the
gauge-field approach. Using a gauge transformation
the Hamiltonian in Eq.~\eqref{eq_hamil} may be rewritten as follows
(we set $\mathcal{R}=1$ and $\phi=0$ in
Eq.~\eqref{eq_spin_spiral})~\cite{intrinsic_torque_no_soi,chigyromag,PhysRevB.18.5073}:
\bege
H(\vn{r})=H_{0}(\vn{r})-\vn{m}\cdot\hat{\vn{M}}^{\rm eff}\Omega^{\rm
  xc}(\vn{r})
+e \vn{A}^{\rm eff} \cdot  \vn{v},
\ee
where $\vn{v}$ is the velocity operator,
\bege
\vn{A}^{\rm eff}=-\frac{i\hbar}{e}U^{\dagger}(\vn{r})
\frac{\partial U(\vn{r})}{\partial \vn{r}}
\ee
is an effective vector potential,
and
\bege\label{eq_gauge_matrix}
U(\vn{r})=
\left(
\begin{array}{cc}
i
\cos\frac{\vn{q}\cdot \vn{r}}{2} +\sin\frac{\vn{q}\cdot \vn{r}}{2} &0 \\
0 &\sin\frac{\vn{q}\cdot \vn{r}}{2}-i\cos\frac{\vn{q}\cdot \vn{r}}{2}
\end{array}
\right)
\ee
transforms the non-collinear magnetization of the spin spiral 
into the collinear magnetization
\bege
\hat{\vn{M}}^{\rm eff}=
\left(-\sin\theta,0,\cos\theta
\right)^{\rm T}.
\ee
Explicitly, the effective vector potential is given by
\bege
\vn{A}^{\rm eff}=\frac{\hbar}{e}
\left(
\begin{array}{cc}
-\frac{\vn{q}}{2}
&0 \\
0 &\frac{\vn{q}}{2}
\end{array}
\right).
\ee

We denote the two eigenstates of the
matrix
\bege
-\vn{m}\cdot\hat{\vn{M}}^{\rm eff}=\mu_{\rm B}
\left(
\begin{array}{cc}
\cos\theta
&- \sin\theta \\
-\sin\theta &-\cos\theta
\end{array}
\right)
\ee
by $|\uparrow\rangle$
and $|\downarrow\rangle$.
Since $\hat{\vn{M}}^{\rm eff}$ lies in the $xz$ plane,
we have $\langle\uparrow| \sigma_{y} |\uparrow\rangle$=0.
A nonzero expectation value $\langle \sigma_{y} \rangle$
corresponds to a torque. Such a non-zero expectation value
may arise from the perturbation by the effective vector 
potential:
\bege\label{eq_perturbi_gauge}
\begin{aligned}
\frac{
\langle\uparrow| \sigma_{y} |\downarrow\rangle
\langle\downarrow| e\vn{A}^{\rm eff}\cdot \vn{v} |\uparrow\rangle}
{\mathcal{E}_{\uparrow}-\mathcal{E}_{\downarrow}}=-
\frac{\sin\theta}{2}
\frac{\langle\downarrow| \hbar \vn{q}\cdot\vn{v} |\uparrow\rangle}{\mathcal{E}_{\uparrow}-\mathcal{E}_{\downarrow}}.
\end{aligned}
\ee

In centrosymmetric systems response coefficients that involve one
torque operator combined with an odd number of velocity operators vanish,
because the torque operator is parity-even, while the velocity
operator is parity-odd. Since Eq.~\eqref{eq_perturbi_gauge} contains
only a single velocity operator, the final expression has to include
one
more matrix element of $e \vn{A}^{\rm eff}\cdot \vn{v}$.
This matrix element is given by
\bege\label{eq_effvecpot_expect}
\langle\uparrow| e \vn{A}^{\rm eff}\cdot \vn{v} |\uparrow\rangle=\hbar\frac{\cos\theta}{2}
\langle\uparrow| \vn{q}\cdot \vn{v} |\uparrow\rangle.
\ee
Multiplication of Eq.~\eqref{eq_perturbi_gauge} and Eq.~\eqref{eq_effvecpot_expect}
shows that the dependence on $q$ and $\theta$ is given by
\bege
\propto q^2 \sin(\theta)\cos(\theta)\propto q^2 \sin(2\theta)
\ee
in agreement with the findings in sections Sec.~\ref{sec_symmetry} and Sec.~\ref{sec_gradi}. 

\subsection{Computational formalism}
\label{sec_formali}
While the laser-induced adiabatic torque
does not require SOI, the laser-induced non-adiabatic
torque is zero in the full calculation when SOI is not included.
By full calculation we mean one that considers
both intrinsic and extrinsic contributions.
This follows from angular momentum conservation,
which is satisfied by the full calculation, provided 
a conserving approximation~\cite{turek_conserving} is used.

In this work, we compute only the intrinsic contribution, which is
nonzero even without SOI. In order to justify this
approximation, we briefly recall the theory of the
current-induced non-adiabatic torque, which makes use
of similar approximations:
The current-induced non-adiabatic torque 
vanishes in the absence of SOI when a
conserving approximation is used.
Extrinsic contributions from scattering need to 
be added to the intrinsic contribution in order to
obtain a conserving 
approximation~\cite{spin_torques_kohno_tatara_shibata,functional_keldysh_spin_torques}.
Therefore, the intrinsic contribution alone does 
not vanish in calculations without SOI.
However, it has been argued that while SOI
is crucial for a non-zero non-adiabatic torque,
it does not strongly affect the magnitude of the
non-adiabatic torque. 
Therefore, calculating the intrinsic non-adiabatic
torque without including SOI may be useful provided
there is a mechanism for angular momentum transfer to
the lattice in the real system that one wishes 
to describe~\cite{intrinsic_torque_no_soi}.

Since vertex corrections are computationally expensive
and numerically tractable expressions for the vertex corrections
to the laser-induced torques have not been derived yet, we
consider in this work only the intrinsic laser-induced torques
without SOI.
Not including SOI in the calculation allows us to obtain
the electronic structure of spin spirals computationally
efficiently based on the generalized Bloch 
theorem~\cite{noco_flapw}.
In order to compute the laser-induced torques we
employ the same equations as those used previously for collinear 
ferromagnets with SOI~\cite{lasintor}.

When torques are induced by femtosecond laser pulses
the torques appear retarded relative to the pulses.
Retardation times between 330fs and 3ps have been
reported~\cite{Huisman_2016,Rabi_CoFeB}.
The expressions that we derived in Ref.~\cite{lasintor}
and that we use in this work were derived under the
assumption of a continuous laser beam rather than a pulse.
However, comparison between the experimental assessment
of the torques induced by fs laser pulses in Ref.~\cite{Huisman_2016}
and our theory~\cite{lasintor} showed good agreement
in the magnitude of the torques. Therefore, we leave
the investigation of retardation effects for future work
and assess the laser-induced torques assuming a continuous
laser beam in this paper.

\section{Results}
\label{sec_results}
\subsection{Computational details}
We obtain the electronic structure of 
bcc Fe, hcp Co, and L$_{1}$0 FePt
selfconsistently using the DFT program
{\tt FLEUR}~\cite{fleurcode}.
The lattice parameters are
$a=5.4235a_{0}$ (Fe),
$a=4.739a_{0}$, $c=7.693a_{0}$ (Co),
and
$a=5.1445a_{0}$, $c=7.1489a_{0}$ (FePt), where
$a_{0}$ is Bohr's radius.
We apply the generalized Bloch theorem
to treat the spin-spiral states\cite{noco_flapw}.
In order to evaluate the laser-induced
torques we take the expressions given
in Ref.~\cite{lasintor} and we make use
of Wannier interpolation~\cite{properties_from_wannier_interpolation}
for computational speed-up.
For this purpose we disentangle
18 maximally localized Wannier functions per
transition metal atom, where we employ
our interface~\cite{WannierPaper} between {\tt FLEUR}
and the {\tt Wannier90} program~\cite{wannier90communitycode}.
The Green's function formalism that we developed in
Ref.~\cite{lasintor} allows us to control disorder through
a quasiparticle broadening parameter $\Gamma$, which we  
set to $\Gamma=25$meV in this paper.
We set the intensity of the laser beam to $I=10{\rm GW}/{\rm cm}^2$
and the photon energy to 1.55~eV.

\subsection{bcc Fe}
\begin{figure}
  \includegraphics[height=0.36\linewidth]{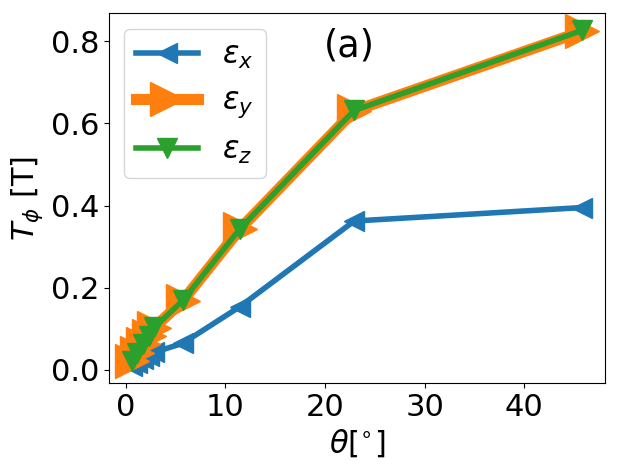}
  \includegraphics[height=0.36\linewidth]{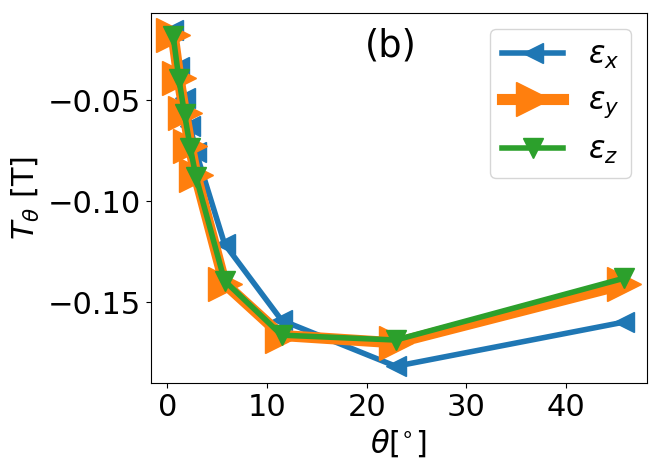}
\caption{\label{fe_adia_vs_theta}
  Laser-induced torques in Fe
  vs.\ cone angle $\theta$ when $\vn{q}=(0,-0.023,0.023)^{\rm T}/a_{0}$.
  (a) Adiabatic torque. (b) Non-adiabatic torque.
}
\end{figure}

In Fig.~\ref{fe_adia_vs_theta}
we show the laser-induced torques in Fe
as a function of cone angle $\theta$ for $q$-vector
$\vn{q}=0.02\vn{b}_{3}=(0,-0.023,0.023)^{\rm T}/a_{0}$ and the three linear polarizations
$\vn{\epsilon}_{x}=(1,0,0)$, $\vn{\epsilon}_{y}=(0,1,0)$,
and $\vn{\epsilon}_{z}=(0,0,1)$. Here, $\vn{b}_{3}$ is the reciprocal
lattice vector 
\bege
\vn{b}_{3}=2\pi\frac{\vn{a}_1\times\vn{a}_2}{(\vn{a}_1\times\vn{a}_2)\cdot\vn{a}_3},
\ee
where $\vn{a}_{i}$ are the primitive lattice vectors.
The polarizations $\vn{\epsilon}_{y}$ and $\vn{\epsilon}_{z}$
yield the same torques, while $\vn{\epsilon}_{x}$ yields different torques,
because the chosen $q$-vector $\vn{q}=(0,-0.023,0.023)^{\rm T}/a_{0}$
lies in the $yz$ plane.
For small cone angle $\theta$ both the adiabatic and the non-adiabatic
torques increase in magnitude with increasing $\theta$. However, the
slope decreases with increasing $\theta$ and in the case of the
non-adiabatic torque the magnitude decreases after reaching a
maximum close to $20^{\circ}$.
This shows that the dependence on $\theta$ is not perfectly described by a 
simple $\sin(2\theta)$, which describes only the leading order in the
expansion with respect to $\theta$ (see Sec.~\eqref{sec_symmetry}), and
higher-order terms in the angular expansion are important.

\begin{figure}
  \includegraphics[height=0.36\linewidth]{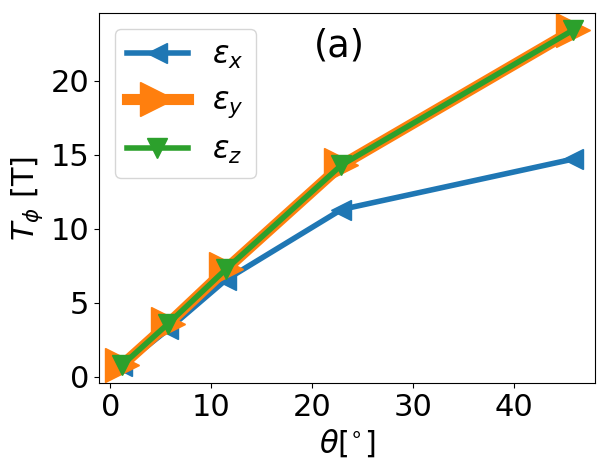}
  \includegraphics[height=0.36\linewidth]{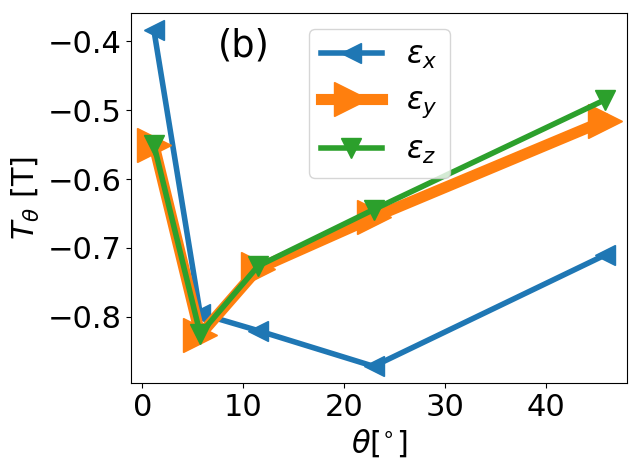}
\caption{\label{fe_adia_vs_theta_q0p1}
  Laser-induced torques in Fe
  vs.\ cone angle $\theta$ when $\vn{q}=(0,-0.115,0.115)^{\rm T}/a_{0}$.
  (a) Adiabatic torque. (b) Non-adiabatic torque.
}
\end{figure}

In Fig.~\ref{fe_adia_vs_theta_q0p1}
we show again the laser-induced torques in Fe
as a function of cone angle $\theta$ but now 
for  a larger $q$-vector of
$\vn{q}=0.1\vn{b}_3=(0,-0.115,0.115)^{\rm T}/a_{0}$.
In this case the nonadiabatic torque reaches a maximum already
close to 5$^{\circ}$ for the polarizations $\vn{\epsilon}_{y}$
and $\vn{\epsilon}_{z}$. Compared to Fig.~\ref{fe_adia_vs_theta}
both the adiabatic and the non-adiabatic torque are larger due to
the larger $q$.
In Ref.~\cite{lasintor} we computed the IFE and OSTT in bcc Fe and
obtained 15~mT and 33~mT, respectively, at the same laser intensity and
quasiparticle broadening as in this paper. In comparison, the
adiabatic torques
in Fig.~\ref{fe_adia_vs_theta_q0p1}(a) are larger by almost three
orders of magnitude.
We attribute these large torques to the strong SOI-like interaction
from
the non-collinearity discussed in Sec.~\ref{sec_gauge_field_approach}.

\begin{figure}
  \includegraphics[height=0.36\linewidth]{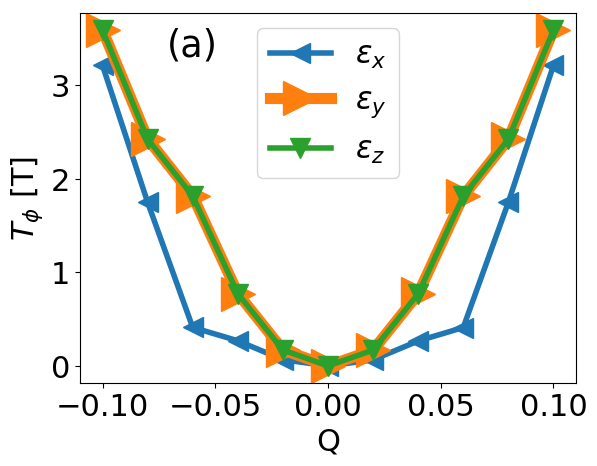}
  \includegraphics[height=0.36\linewidth]{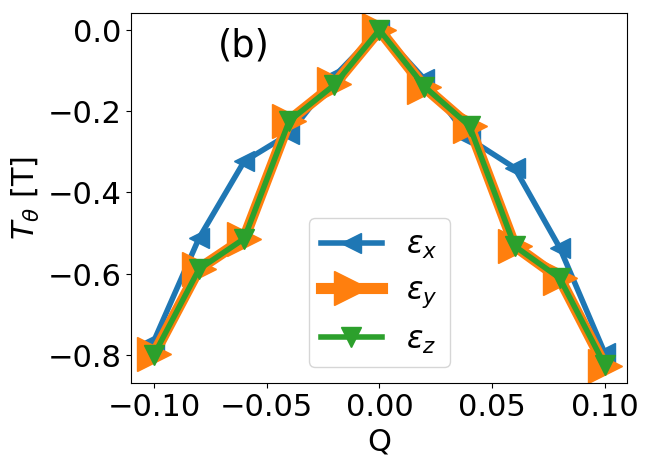}
\caption{\label{fe_adia_vs_q}
  Laser-induced torques in Fe
  vs.\ $q$-vector $\vn{q}=(0,-1.1585Q,1.1585Q)^{\rm T}/a_{0}$ when
  $\theta=5.7^{\circ}$.
  (a) Adiabatic torque. (b) Non-adiabatic torque.
}
\end{figure}

In order to investigate the $q$-dependence in more detail  we show in
Fig.~\ref{fe_adia_vs_q}
the laser-induced torques in Fe
as a function of $q$-vector $\vn{q}=Q\vn{b}_{3}=(0,-1.1585Q,1.1585Q)^{\rm T}/a_{0}$ when
  $\theta=5.7^{\circ}$. The torques increase monotonously with $Q$
and they are even in $Q$. For small $Q$ the nonadiabatic torque
behaves like $\propto |Q|$, while the adiabatic torque behaves like
$\propto Q^2$. This behaviour is consistent with the symmetry analysis
in Sec.~\ref{sec_symmetry} predicting the torques to be even in
spin-spiral
wave vector $q$.

\subsection{hcp Co}

\begin{figure}
  \includegraphics[height=0.36\linewidth]{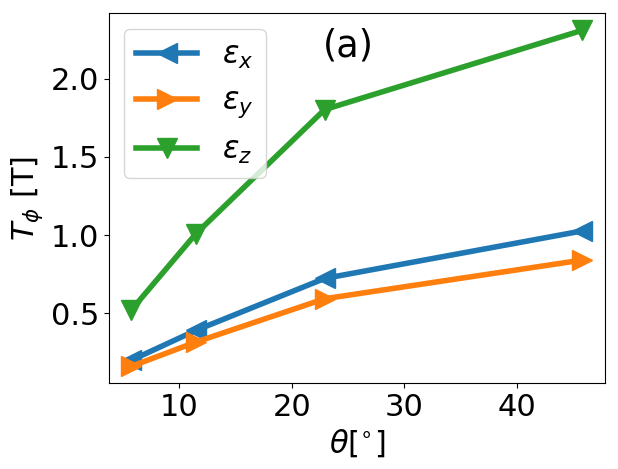}
  \includegraphics[height=0.36\linewidth]{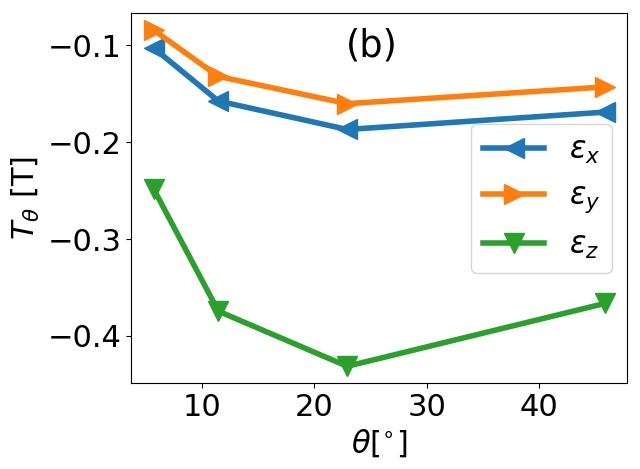}
\caption{\label{co_adia_vs_theta}
  Laser-induced torques in Co
  vs.\ cone angle $\theta$ when $\vn{q}=(0,0,0.016)^{\rm T}/a_{0}$.
  (a) Adiabatic torque. (b) Non-adiabatic torque.
}
\end{figure}

In Fig.~\ref{co_adia_vs_theta}
we show the laser-induced torques in Co
as a function of cone angle $\theta$ for $q$-vector
$\vn{q}=0.02\vn{b}_3=(0,0,0.016)^{\rm T}/a_{0}$.
The in-plane polarizations $\vn{\epsilon}_{x}$ and $\vn{\epsilon}_{y}$
yield very similar torques, because the $q$-vector points in the
$z$ direction and is therefore perpendicular to both polarizations. 
The slight difference between torques
for $\vn{\epsilon}_{x}$ and $\vn{\epsilon}_{y}$
can be explained by considering that the $x$ and $y$ directions
in the hexagonal unit cell are not equivalent. Similar to the case of
bcc Fe shown in Fig.~\ref{fe_adia_vs_theta} the nonadiabatic torque
attains a maximum already close to 20$^{\circ}$ and therefore requires
higher-order terms in the angular expansion beyond the leading order
term $\propto \sin(2\theta)$ for its description.

\begin{figure}
  \includegraphics[height=0.36\linewidth]{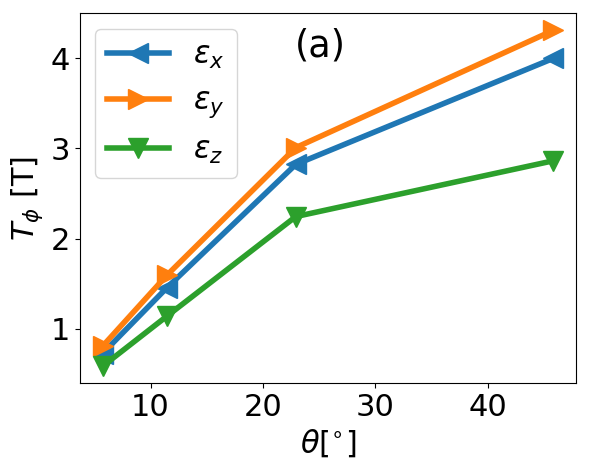}
  \includegraphics[height=0.36\linewidth]{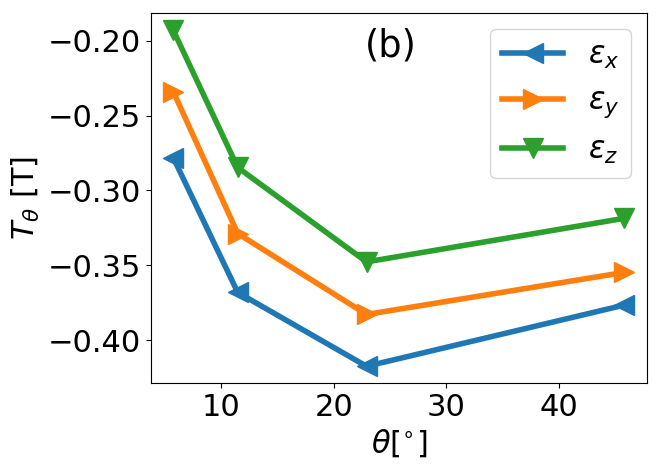}
\caption{\label{co_adia_vs_theta_qx}
  Laser-induced torques in Co
  vs.\ cone angle $\theta$ when $\vn{q}=(0.015,-0.0265,0)^{\rm T}/a_{0}$.
  (a) Adiabatic torque. (b) Non-adiabatic torque.
}
\end{figure}

In Fig.~\ref{co_adia_vs_theta_qx}
we show the laser-induced torques in Co
as a function of cone angle $\theta$ for the $q$-vector
$\vn{q}=0.02\vn{b}_1=(0.015,-0.0265,0)^{\rm T}/a_{0}$, which lies in the
$xy$ plane. The non-adiabatic torque is now different between 
the polarizations $\vn{\epsilon}_{x}$ and $\vn{\epsilon}_{y}$,
because the $q$-vector forms different angles with the
$x$ and $y$ axes.

For the same laser intensity and quasiparticle broadening as used in
this paper we determined the IFE and OSTT in Co in
Ref.~\cite{lasintor}
and obtained 118~mT and 0.229~mT, respectively. In comparison,
the adiabatic torques shown in Fig.~\ref{co_adia_vs_theta}(a)  and in Fig.~\ref{co_adia_vs_theta_qx}(a) 
are orders of magnitude larger.

\subsection{L$_{1}$0 FePt}
\begin{figure}
  \includegraphics[height=0.36\linewidth]{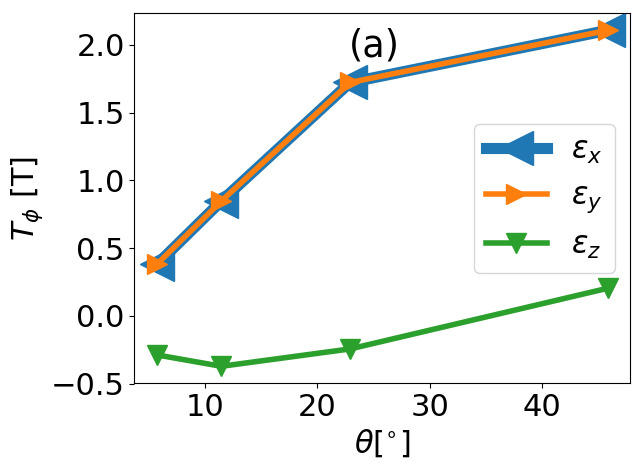}
  \includegraphics[height=0.36\linewidth]{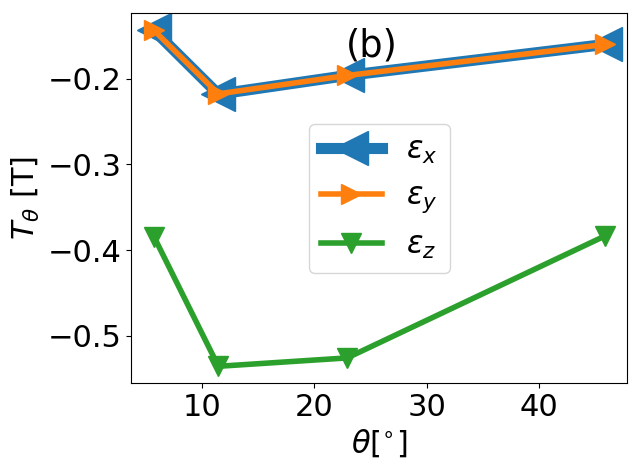}
\caption{\label{fept_adia_vs_theta}
  Laser-induced torques in FePt
  vs.\ cone angle $\theta$ when $\vn{q}=(0,0,0.0176)^{\rm T}/a_{0}$.
  (a) Adiabatic torque. (b) Non-adiabatic torque.
}
\end{figure}

In Fig.~\ref{fept_adia_vs_theta}
we show the laser-induced torques in FePt
as a function of cone angle $\theta$ for $q$-vector
$\vn{q}=0.02\vn{b}_3=(0,0,0.0176)^{\rm T}/a_{0}$.
The torques for the polarizations
$\vn{\epsilon}_{x}$ and $\vn{\epsilon}_{y}$
agree, because the crystal axes $a$ and $b$ are
equivalent and because the $q$ vector is perpendicular to both of them.
At large angles $\theta$ the adiabatic torque for polarization $\vn{\epsilon}_{z}$
is strongly suppressed in Fig.~\ref{fept_adia_vs_theta}(a) due to the
large
anisotropy in the L$_{1}$0 structure.
Similar to the cases of
bcc Fe and
hcp Co shown in Fig.~\ref{fe_adia_vs_theta} 
and Fig.~\ref{co_adia_vs_theta}, respectively,
the nonadiabatic torque
attains a maximum already close to 20$^{\circ}$ and therefore requires
higher-order terms in the angular expansion beyond the leading order
term $\propto \sin(2\theta)$ for its description.

\begin{figure}
  \includegraphics[height=0.36\linewidth]{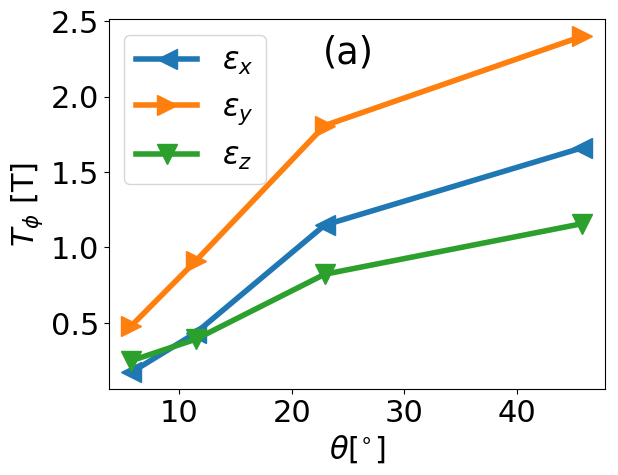}
  \includegraphics[height=0.36\linewidth]{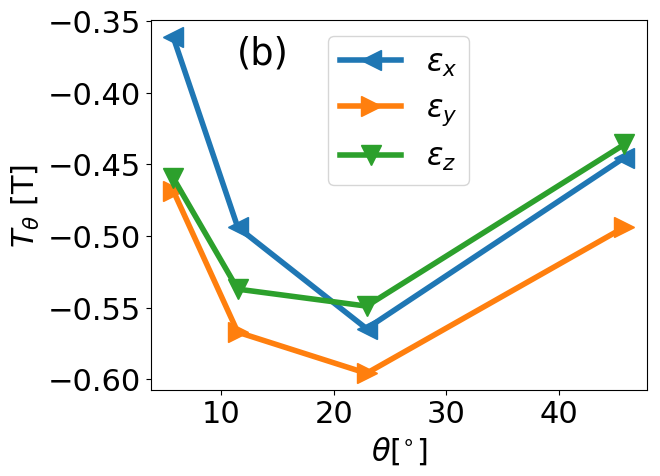}
\caption{\label{fept_adia_vs_theta_qx}
  Laser-induced torques in FePt
  vs.\ cone angle $\theta$ when $\vn{q}=(0.0244,0,0.0)^{\rm T}/a_{0}$.
  (a) Adiabatic torque. (b) Non-adiabatic torque.
}
\end{figure}

In Fig.~\ref{fept_adia_vs_theta_qx}
we show the laser-induced torques in FePt
as a function of cone angle $\theta$ for $q$-vector
$\vn{q}=0.02\vn{b}_1=(0.0244,0,0)^{\rm T}/a_{0}$.
In this case the torques are different for the three
polarizations $\vn{\epsilon}_{x}$, $\vn{\epsilon}_{y}$
and $\vn{\epsilon}_{z}$: The $x$ and $y$ directions
are inequivalent, because $\vn{q}$ points into x direction,
and the $z$ direction is inequivalent to the $y$ direction,
because the $c$-axis is longer than the $b$-axis.

For the same laser intensity and quasiparticle broadening as used in
this paper we determined the IFE and OSTT in FePt in
Ref.~\cite{lasintor}
and obtained 185~mT and 22~mT, respectively. In comparison,
the adiabatic torques shown in Fig.~\ref{fept_adia_vs_theta}(a)  and in Fig.~\ref{fept_adia_vs_theta_qx}(a) 
are more than one order of magnitude larger.

\section{Conclusion}
\label{sec_conclusion}
We investigate laser-induced torques in homogeneous spin spirals
without
spin-orbit interaction (SOI)
using symmetry arguments and first principles calculations.
Symmetry analysis shows that laser-induced torques vanish for flat
spirals -- at the leading order of an angular expansion the dependence
on spiral cone angle is $\propto\sin(2\theta)$ --
and that their dependence on the spin-spiral wave-vector $\vn{q}$ is
even in $q$. Additionally, it shows that laser-induced torques in
homogeneous spin-spirals are not associated with spin currents.
Our first-principles calculations show that the laser-induced torques
in bcc Fe, hcp Co and L$_{1}$0 FePt with an imposed spin-spiral
magnetic
structure may be orders of magnitude larger than those in the
corresponding
magnetically collinear systems with SOI.
This suggests that these torques may play an important role in
ultrafast magnetism phenomena.
Within the frozen-magnon approximation our results may also be 
used to estimate the laser-induced torques on magnons in these
materials.
\section*{Acknowledgments}
We acknowledge financial support from Leibniz Collaborative Excellence project OptiSPIN $-$ Optical Control of Nanoscale Spin Textures, and funding  under SPP 2137 ``Skyrmionics" of the DFG. We gratefully acknowledge financial support from the European Research Council (ERC) under the European Union's Horizon 2020 research and innovation program (Grant No. 856538, project "3D MAGiC”), and ITN Network COMRAD. The work was also supported by the Deutsche Forschungsgemeinschaft (DFG, German Research Foundation) $-$ TRR 173 $-$ 268565370 (project A11), TRR 288 – 422213477 (projects B06).  We  also gratefully acknowledge the J\"ulich Supercomputing Centre and RWTH Aachen University for providing computational resources under project No. jiff40.

\bibliography{lasintospira}

\end{document}